\documentclass[doublecol]{epl2} 
\usepackage{cancel}
\usepackage{mathrsfs}
\usepackage{bbold}
\usepackage{float}
\usepackage{graphicx}
\usepackage{bm}% bold math
\usepackage{amsmath} % \pmb bold special characters
\usepackage{booktabs}
\usepackage{color}

\def\ket#1{\mathinner{|{#1}\rangle}}

\def\<{\langle}
\def\>{\rangle}

\title{Spin texture of an irradiated warped topological insulator surface}
\shorttitle{Spin texture of an irradiated warped topological insulator surface} %Insert here a short version of the title if it exceeds 70 characters

\author{Debabrata Sinha\inst{1,2}}

\institute{                    
  \inst{1} The Institute of Mathematical Sciences, C.I.T. Campus, Chennai 600 113, India\\
  \inst{2} TIFR Centre for Interdisciplinary Sciences, Hyderabad 500075, India
}

\abstract{Topological insulator is a new state of matter which exhibits exotic surface electronic properties. Determining the spin texture of this class of material is of paramount importance for understanding its topological order and can lead to potential applications in spintronics. Here, we have investigated the nature of the surface state of the topological insulator with hexagonal warping subjected to an off-resonant circularly polarized light. The resulting electronic ground state exhibits a novel feature of spin texture breaking the conventional spin-momentum locking present on a topological insulator surface. The observed spin texture is shown to be a consequence of the symmetry group of the underlying crystal. The generalisation of our method to the other 2D graphene-like systems are straightforward. Our calculation charts a simple experimental route for a realisation of the non-trivial spin-textures.}

\begin{document}

\maketitle
\section{Introduction}

Topological Insulators (TIs) are unique phases of matter which are bulk insulators but have topologically protected metallic surface states. The surface of a two dimensional TI contains even number of Dirac points and exhibits various extraordinary properties like robust gapless edge states \cite{Zhang-08,Zhou-08}, quantized hall conductivity \cite{kane-rmp}, topological excitations \cite{Hughes-Nat08} etc. However, surface states of some recently discovered three-dimensional (3D) TIs like Bi$_{2}$Se$_{3}$ and Bi$_{2}$Te$_{3}$ have been shown to have only a single Dirac cone on the surface \cite{Zhang-Nat09,Niu-Nat10,Chadov-NAM10,Hsieh-Nat09,Hsieh-PRL09}. The 3D TIs have also been shown to possess an unconventional electronic phase with spin-momentum locking on the surface of a Dirac cone and have weak anti-localisation effect in its transport properties\cite{Lu-PRL11,Hikami-PTP80,Bergmann-PR84}.The conventional spin-momentum locking is an important property of a gapless 3D TI and has been observed in experiments too\cite{Hsieh-Nat09},\cite{Su-Science11},\cite{Hsieh-Nat08}. The spin-momentum locking naturally leads to current-induced spin polarization on the surface states \cite{OV-PRL10} of a TI and has important implications for spintronics. Time reversal symmetry breaking by ferromagnetic doping on the surface of these class of TIs can break the locking and lead to a non-trivial spin texture \cite{Liu-Nature12},\cite{Qin-Niu-PRL}. The unique spin texture is a possible signature of Dirac-metal to gapped-insulator transition.

The spin-momentum unlocking can also arise naturally for a time reversal symmetric TI with a finite warping term. Recent angle-resolved photoemission (ARPES) experiments showed that $Bi_{2}Te_{3}$ has a rhombohedral structure with significant hexagonal warping \cite{Hsieh-Nat09},\cite{Hsieh-PRL09},\cite{Fu,Chen-Scince10,Li-PRB13,Li-PRB14}. The Fermi surface in the observed band structure develops from a circle to a hexagon to a snowflake like shape with increasing chemical potential. Although the hexagonal warping term costs a finite out-of-plane spin component but the in-plane spin texture remains perpendicular to the momentum direction. A higher order warping term is also possible within the symmetry group of the given crystal. It was shown that the higher order term breaks the spin-momentum locking on the surface of a Dirac cone \cite{Bansil-PRB11}.

Here, we will follow a different path and demonstrate that a non-trivial spin texture also emerges on a hexagonal warped TI surface exposed to an off-resonant light. We will show that the hexagonal warping term plays an important role to modify the Dresselhaus spin orbit coupling on TI surface. The resultant electronic spin texture becomes non-orthogonal with the momentum. We attempt to answer the following questions: What are the fundamental differences between the observed novel spin-texture of a gapped and gapeless TI with finite warping? What are  the differences in spin-texture for a magnetically doped TI and a time reversal breaking warped TI? We show that the fundamental differences lies into the symmetry group of the underlying crystal. % We will show that the differences of the resulting spin-texture can be elegantly explained using symmetry arguments.} 

\section{Model and Floquet Theory}
We start our discussion by introducing Hamiltonian which describes the surface states of a TI with finite hexagonal warping \cite{Fu},
\begin{eqnarray}
\mathcal{H}_{0}(\vec{k})=\hbar v_{k}(k_{x}\sigma_{y}-k_{y}\sigma_{x})+\frac{\lambda}{2}(k^3_{+}+k^3_{-})\sigma_{z}
\label{Hamil}
\end{eqnarray}
The first term in the Hamiltonian describes spin-orbit locking with $v_{k}$ being the Fermi velocity. In general, the Fermi velocity may contain second order correction term i.e., $v_{k}=v_{F}(1+\gamma k^2)$. However, here we only consider the case of constant velocity and for the rest of the paper, $v_{k}=v$. $\sigma_{x}$,$\sigma_{y}$,$\sigma_{z}$ are the Pauli matrices. $k_{\pm}=k_{x}\pm ik_{y}$ where $k_{x}$,$k_{y}$ are the momentum along the $x$ and $y$ axis, respectively. The last term of the Hamiltonian in Eq.(\ref{Hamil}) describes cubic spin-orbit coupling at the surface of rhombohedral crystal systems and responsible for hexagonal distortion at the Fermi surface. It breaks $U(1)$ symmetry in the system and is only invariant under threefold rotation and mirror transformation along the $x$ axis ($C_{3v})$, where $x$ is along $\Gamma-K$ direction\cite{Fu}. Note that higher order terms that preserve the $C_{3v}$ invariance are possible and their inclusion is not going to change any of the conclusions of this paper. We note that though the Hamiltonian explicitly break $U(1)$ invariance, there is, at this order in $k$, no term that breaks the locking of spin and momentum since there is no term that couples $\sigma_{x,y}$ to higher powers of $k_\pm$. Thus, if the Hamiltonian is only expanded till this order, the spin remains locked to the momentum in the time-reversal symmetric problem. This is, however, not a consequence of any symmetry of the problem and the spin-momentum locking can be broken by extending the Hamiltonian to include the next order term allowed by the $C_{3v}$ symmetry of the system\cite{Bansil-PRB11}. %We follow a different path and demonstrate that non-trivial spin textures emerge if time-reversal symmetry is broken. We will show that the spin structure we obtain is distinctly different from the one obtained when the spin-momentum locking is broken by including higher order terms in the Hamiltonian.}

We now consider a time-reversal-symmetry-breaking contribution arising from the uniform irradiation of the TI surface by an optically polarised light beam. The electric field of the polarized light is given by, $E(t)=E_{0}(\cos(\omega t),-\sin(\omega t))$, where $E_{0}$ and $\omega$ are the amplitude and frequency of the optical field, respectively. The corresponding vector potential is $\vec{A}(t)=A_{0}(\sin(\omega t),\cos(\omega t))$ with $A_{0}=-E_{0}/ \omega$. The optical field enters into the Hamiltonian via the Peierls substitution which replaces the momentum of electron by $\hbar k_{i} \rightarrow \hbar k_{i}+eA_{i}$. The vector potential of the polarized optical field is periodic in time i.e., $\vec{A}(t+T)=\vec{A}(t)$ with $T=2\pi/ \omega$ and so the Hamiltonian also becomes time periodic. In order to find the eigenstates of the time-periodic Hamiltonian we will take advantage of the Floquet formalism. Recently, several authors have used Floquet formalism in the context of topological phenomena of TI and graphene\cite{Demler-PRB11,Cayssol-PSS13,Podo-PRB13,Lago-PRA15,Calvo-PRB15,Jun-PRL10,Dora-PRL12,Lovey-ar}. For completeness, we briefly discuss the basic idea of Floquet formalism in the Appendix. We further consider that the frequency of the optical field is off-resonant and does not cause any electronic transition. This can be achieved if the photon energy of the polarized light is higher than the bandwidth of our system i.e., the frequency lies in the soft-x-ray regime ($10^{15}$Hz). In the off-resonant condition, it is sufficient to consider two low-order process in the Floquet regime, describing a single virtual photon emission and absorption. This off-resonant formalism has been successfully applied to graphene, TI and Weyl semimetal \cite{Zhai-PRB14,Zhou-PRB15,Tahir-PRB,weyl-semi} and the results also agree with experiments\cite{Wang-Science13},\cite{Rechtsman-Nat13}. We adapt this here to study spin-texture of topological insulators in an optical field. Also, there is another type of light called on-resonant light whose frequency ranges from far infrared ($10^{12}$Hz) to visible light ($10^{14}$Hz). The on-resonant light can induce interband and intraband electron transition which we are not taking consideration here.

Under irradiation, the time dependent Hamiltonian is given by,
\begin{eqnarray}
\mathcal{H}(\vec{k},t)=\mathcal{H}_{0}(\vec{k})+\mathcal{V}(t)
\end{eqnarray}
where $\mathcal{H}_{0}$ is given in Eq.(\ref{Hamil}) and $\mathcal{V}(t)=ev[A_{x}\sigma_{y}-A_{y}\sigma_{x}]+\frac{3e\lambda}{2\hbar}[k^2_{+}A_{+}+k^2_{-}A_{-}]\sigma_{z}+\mathcal{O}(A^2,A^3)$ with $A_{\pm}=A_{x}\pm  iA_{y}$. The effective Hamiltonian is approximately expressed as,
\begin{eqnarray}
\mathcal{H}_{eff}=\mathcal{H}_{0}+\frac{[V_{-1},V_{+1}]}{\hbar \omega}
\label{eff-hamiltonian}
\end{eqnarray}
with $V_{-1}=i\alpha(i\sigma_{x}-\sigma_{y})+i\beta[2ik_{x}k_{y}-(k^{2}_{x}-k^{2}_{y})]\sigma_{z}$ and $V_{+1}=V^{\dagger}_{-1}$. Where $\alpha=\frac{evA_{0}}{2}$ and $\beta=\frac{3e\lambda A_{0}}{2\hbar}$. Using the form of $V_{-1}$ and $V_{+1}$, Eq.(\ref{eff-hamiltonian}) becomes,

\begin{eqnarray}
\mathcal{H}_{eff}&=&\hbar v[(k_{x}+\mathcal{K}_{1}(k_{x},k_{y}))\sigma_{y}-(k_{y}+\mathcal{K}_{2}(k_{x},k_{y}))\sigma_{x}]\nonumber\\&&+\frac{\lambda}{2}(k^3_{+}+k^{3}_{-})\sigma_{z}+\Delta_{\omega} \sigma_{z}
\label{Eff-Hamil}
\end{eqnarray}
where $\mathcal{K}_{1}(k_{x},k_{y})=-\frac{4\alpha \beta}{\hbar^2\omega v}(k^2_{x}-k^2_{y})$, $\mathcal{K}_{2}(k_{x},k_{y})=\frac{8\alpha \beta}{\hbar^2\omega v}k_{x}k_{y}$ and $\Delta_{\omega}=\frac{4\alpha^2}{\hbar \omega}$.
From the above Hamiltonian, it is evident that the influence of the off-resonant optical field on the band structure of the system is two fold: it renormalizes the spin-orbit coupling strength of TI surface and introduces a gap. Note that in the absence of the warping term both $\mathcal{K}_{1}$ and $\mathcal{K}_{2}$ are zero but $\Delta_{\omega}\neq 0$. Such kind of Floquet systems have been studied in the recent past in the context of graphene and TI \cite{Xin-CHIN14,Zhai-PRB14,Zhou-PRB15,Lovey-ar,Ezawa-PRL13,Tahir-PRB}. The warping term here modifies the spin-orbit coupling on the TI surface. The Hamiltonian in Eq.(\ref{Eff-Hamil}) breaks the time reversal symmetry. Also, the term $\mathcal{K}_{1}\sigma_{y}-\mathcal{K}_{2}\sigma_{x}$ in Eq.(\ref{Eff-Hamil}) breaks mirror symmetry that is present in Eq.(\ref{Hamil}). The symmetries of the Floquet Hamiltonian are important in understanding the emergent spin texture phenomena.

The matrix form of Eq.(\ref{Eff-Hamil}) is given by,
\begin{eqnarray}
\mathcal{H}_{eff}=\begin{pmatrix}
\Delta(k,\theta)& \hbar v(-ik_{-}+iak^2_{+})\\
\hbar v(ik_{+}-iak^2_{-})& -\Delta(k,\theta)
\end{pmatrix}
\end{eqnarray}
where $\Delta(\vec{k},\theta)=\lambda k^3\cos(3\theta)+\Delta_{\omega}=\lambda (k^3_{x}-3k_{x}k^2_{y})+\Delta_{\omega}$, with $\theta=\tan^{-1}(k_{y}/k_{x})$ and $a=\frac{4\alpha\beta}{\hbar^{2}\omega v}$. The energy eigenvalues are given by,
\begin{eqnarray}
\mathcal{E}(\vec{k})=s\sqrt{(\Delta(\vec{k},\theta))^{2}+\hbar^{2}v^{2}(k^2+a^{2}k^{4}-2ak^3\cos(3\theta))}
\end{eqnarray}
where $s=\pm$, defines conduction and valance band respectively. We see that the band structure is three-fold symmetric under $\theta\rightarrow \theta \pm 2\pi/3$. This is different from the symmetry of the band structure of the time-reversal symmetric Hamiltonian, which, because of time-reversal symmetry, is six-fold symmetric. The wave function $u(\vec{k},s)$ is defined by the equation $\mathcal{H}(\vec{k})_{eff}u(\vec{k},s)=\mathcal{E}(\vec{k})u(\vec{k},s)$ and is given by,
\begin{eqnarray}
u(k,s)=\mathcal{C}_{s}\begin{pmatrix}
1\\
\frac{\hbar v(ik_{+}-iak^2_{-})}{\mathcal{E}_{s}+\Delta(k,\theta)}
\end{pmatrix}
\label{wavef}
\end{eqnarray}
where $\mathcal{C}_{s}$ is the normalization parameter and 
\begin{eqnarray}
\mathcal{C}^{2}_{s}=[1+\frac{\hbar^2v^2(k^2+a^2k^4-2ak^3\cos(3\theta))}{(\mathcal{E}_{s}+\Delta(k,\theta))^2}]^{-1}
\end{eqnarray}
For rest of the paper, we have used $\hbar \omega=8 eV$. We choose $evA_{0}=0.5 eV$ and $ev A_{0}=0.9 eV$. The corresponding values of $a=0.17 nm$, $\Delta_{\omega}=0.03 eV$ ($evA_{0}=0.5 eV$) and $a=0.55 nm$, $\Delta_{\omega}=0.1 eV$ ($evA_{0}=0.9 eV$). With these values, $\hbar \omega \gg evA_{0}$ i.e. off-resonant conditions are satisfied. We will present the results here for $s=+1$ i.e., for the conduction bands.

\begin{figure}
\center
\rotatebox{0}{\includegraphics[width=2.0in]{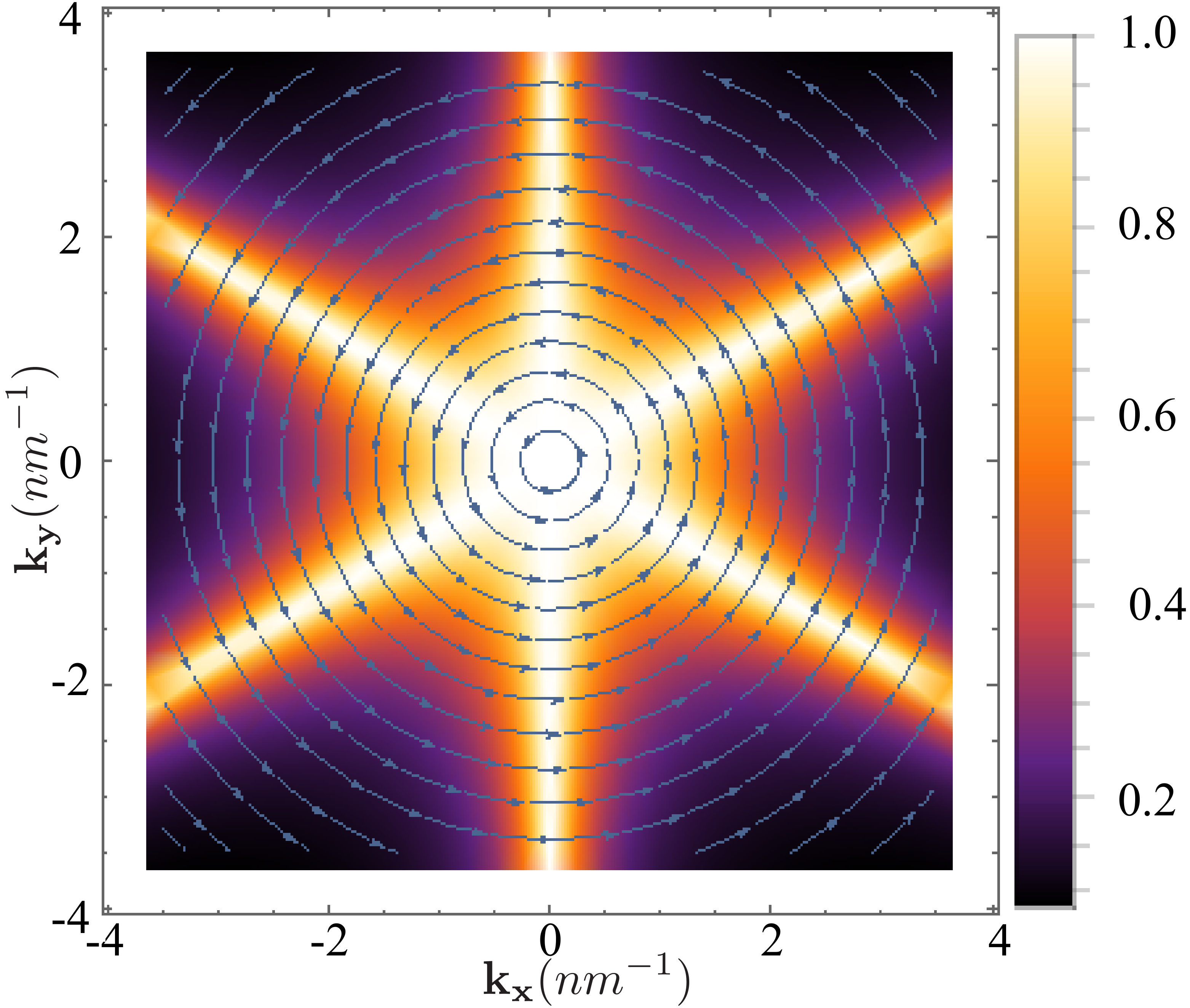}}
\caption{Spin Texture ($S_{x},S_{y}$) in the $k_{x}-k_{y}$ plane of the surface state Brillouin zone. The unit of momentum is nm$^{-1}$ (gapless). The gap $\Delta_{\omega}=0$ eV, the warping $\lambda=0.2$ eV nm$^{3}$ and $a=0$ (no optical polarized field). The colour shows the in-plane spin density.}
\label{spin-plot}
\end{figure}
\begin{figure}
\rotatebox{0}{\includegraphics[width=3.2in]{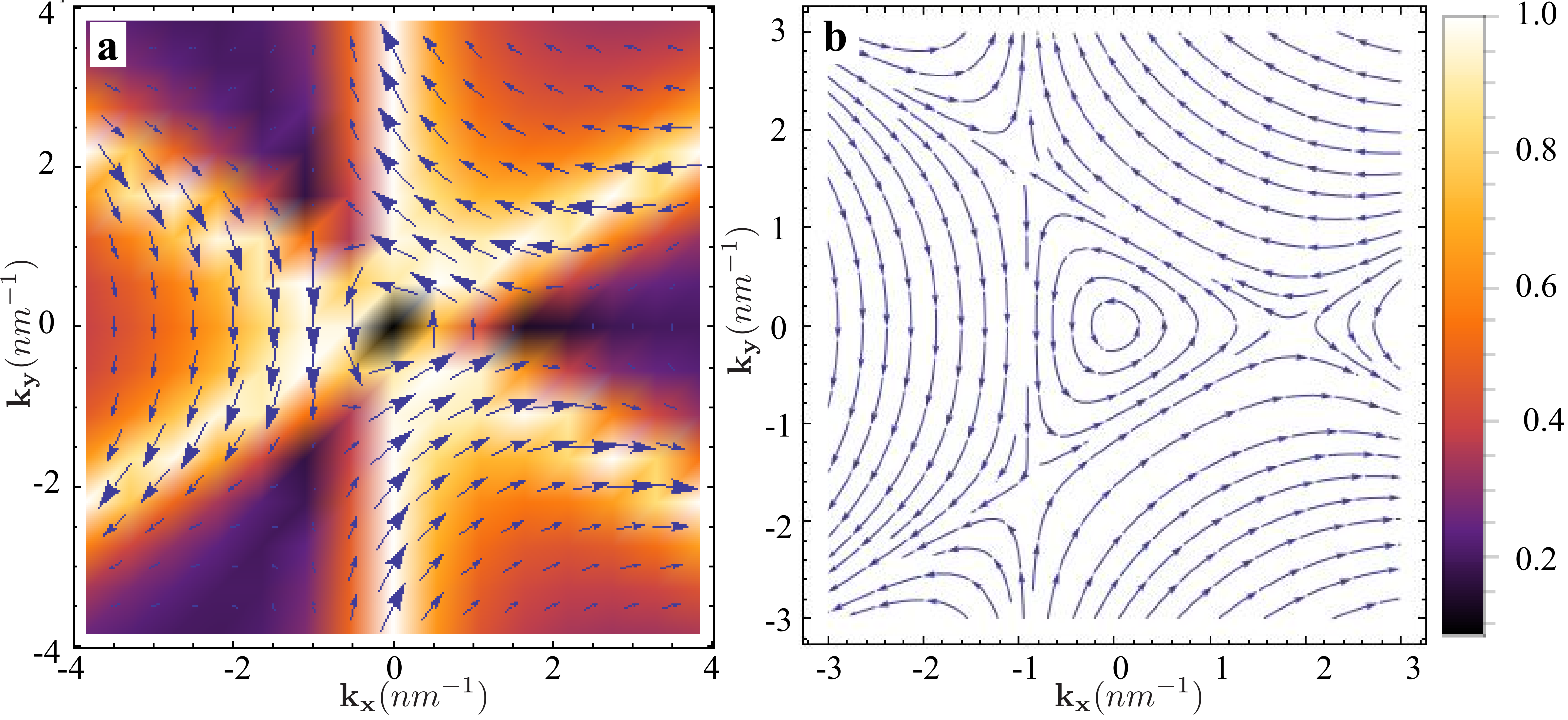}}
%\rotatebox{0}{\includegraphics[width=1.6in]{stream.pdf}}
\caption{(a)Spin texture in the $k_{x}-k_{y}$ plane of a warped Floquet TI surface (gapped). The parameters are $\Delta_{\omega}=0.1$ eV, $\lambda=0.2$ eV nm$^{3}$ and $a=0.55$ nm.(b) Corresponding stream line structure of the spin texture. The resultant spin is not perpendicular with the direction of the momentum.}
\label{spin3l-hg}
\end{figure}

\section{Spin Texture}
Using the eigenfunctions in Eq.(\ref{wavef}) of effective Hamiltonian, one can calculate the average value of spin \cite{Li-PRB14,berche-epj}. We compute the average value of the electron spin components $S_{x}, S_{y}$(in-plane) and $S_{z}$(out-of-plane) and are given by, 
\begin{eqnarray}
\label{in-planex}
S_{x}&=&\frac{\hbar}{2}\mathcal{C}_{s}^2\frac{\hbar v[-4ak_{x}k_{y}-2k_{y}]}{\mathcal{E}_{s}(k,\theta)+\Delta(k,\theta)}\\
%\label{in-planex}
%\end{eqnarray}
%\begin{eqnarray}
\label{in-planey}
S_{y}&=&\frac{\hbar}{2}\mathcal{C}_{s}^2\frac{\hbar v[-2a(k^2_{x}-k^2_{y})+2k_{x}]}{\mathcal{E}_{s}(k,\theta)+\Delta(k,\theta)}\\
%\label{in-planey}
%\end{eqnarray}
%\begin{eqnarray}
S_{z}&=&\frac{\hbar}{2}\mathcal{C}_{s}^2[1-\frac{\hbar^{2}v^2(k^2+a^2k^4-2ak^3\cos(3\theta))}{(\mathcal{E}_{s}(k,\theta)+\Delta(k,\theta))^2}]
\label{out-of-plane}
\end{eqnarray}
In the absence of the polarized optical field ($a=0$ and $\Delta_{\omega}=0$), the results are valid for a gapless TI with finite hexagonal warping. Fig.(\ref{spin-plot}) shows the spin texture on the surface of the Dirac cone of a normal topological insulator with hexagonal wraping. The arrows denote the direction of the spin. The in-plane components of the spin remain locked with the perpendicular component of the momentum on the surface of a Dirac cone, as discussed in \cite{Li-PRB14},\cite{Bansil-PRB11}. The off-resonant radiation modifies the spin-orbit coupling and introduces a gap on the TI surface (see Eq.(\ref{Eff-Hamil})). The spin-momentum locking is no longer present now, which is clear from Eq.(\ref{in-planex}) and Eq.(\ref{in-planey}). Fig.(\ref{spin3l-hg}) shows the in-plane spin texture of a Floquet topological insulator surface with a finite hexagonal warping. The spin texture pattern is symmetric under three-fold rotation $C_{3v}$ which follows from symmetry of the time reversal breaking Hamiltonian in Eq.(\ref{Eff-Hamil}). The higher order warping term as in Ref \cite{Bansil-PRB11} can be taken into account in our model. But from symmetry considerations, it will not change the spin texture pattern discussed here.  Although, the spin texture in Ref.\cite{Bansil-PRB11} is quite similar to Fig.(\ref{spin3l-hg}), the fundamental difference appears in the angle of deviation which we will discuss later in the paper. Another way to observe a non-trivial spin texture of a TI is by breaking time reversal symmetry with a magnetic doping. The magnetic polarisation breaks the residual rotational symmetry of a TI and creates a novel spin texture phenomena\cite{Qin-Niu-PRL}.

\begin{figure}
\begin{center}
\rotatebox{0}{\includegraphics[width=0.95\columnwidth]{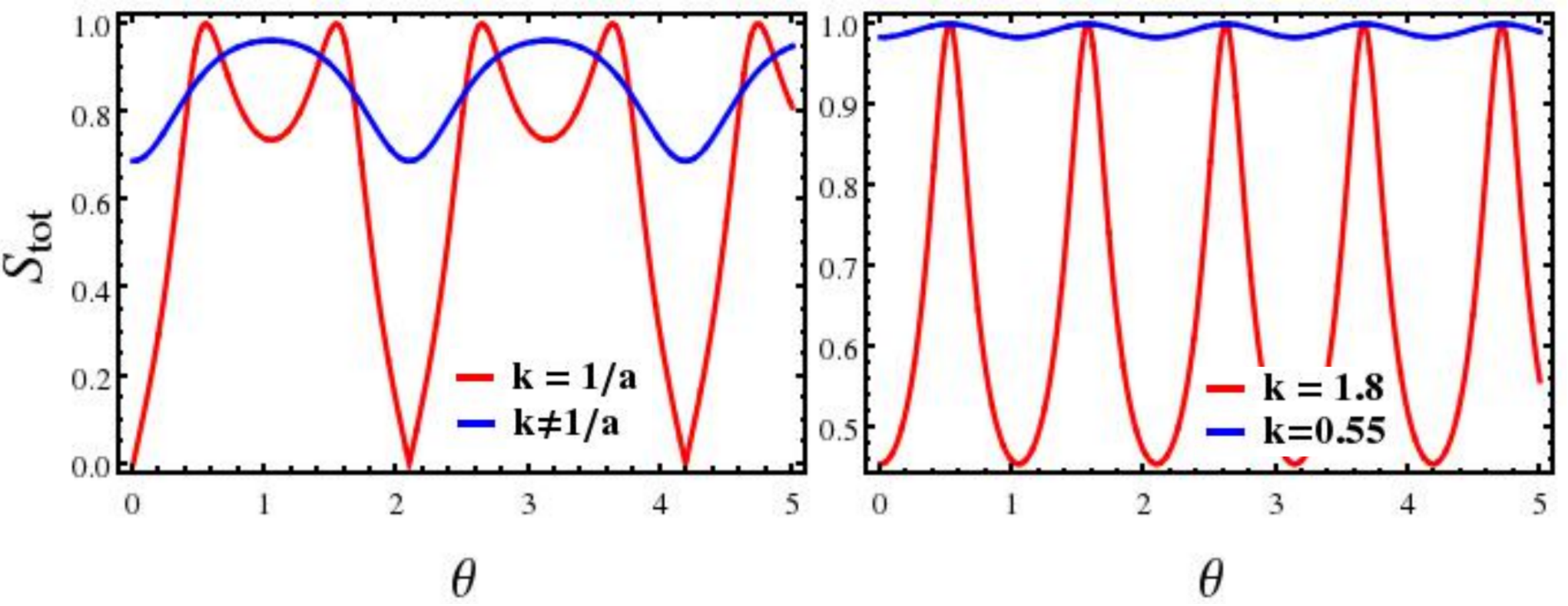}}
%\rotatebox{0}{\includegraphics[width=1.7in]{spin-new-n.pdf}}
\caption{Total spin desity as a function of $\theta$. Left panel shows the results for Floquet system ($a \neq 0$)and right panel show the results in the absence of polarized field ($a=0$ and $\Delta_{\omega}=0$). In left panel, for $k=\frac{1}{a}$, $S_{tot}$ has minimum zero value when $\theta_{min}=2n\pi/3$ ($n \in \mathbb{Z}$). For $k\neq \frac{1}{a}$, $S_{tot}$ has non-zero minimum for $\theta_{min}$. In right panel, $S_{tot}$ is almost constant for lower vales of $k$. For higher values of $k$, $S_{tot}$ oscillates due to significant contributions of the warping term.}
\label{spin-den}
\end{center}
\end{figure}
\begin{figure}
\rotatebox{0}{\includegraphics[width=1.72in]{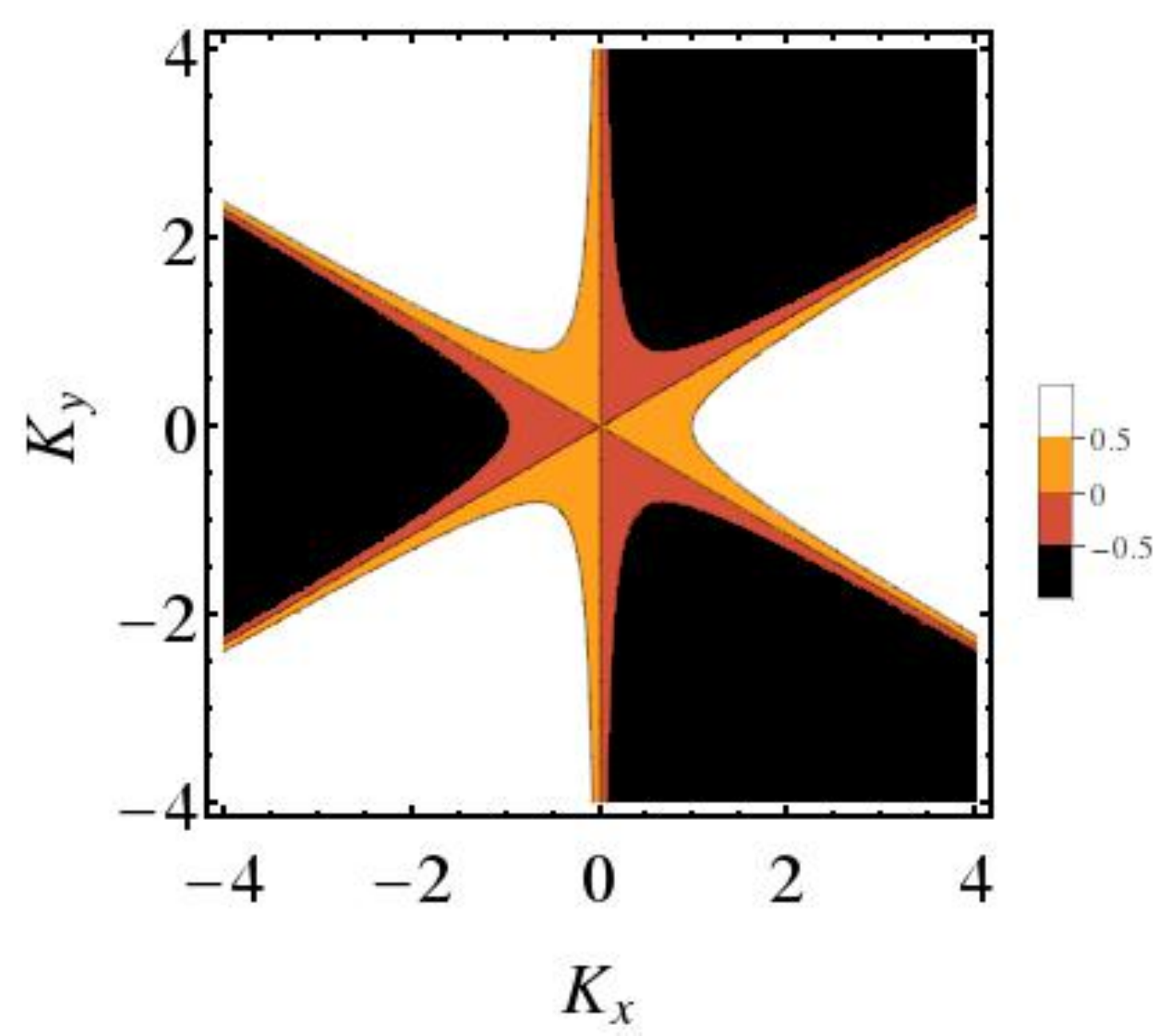}}
\rotatebox{0}{\includegraphics[width=1.72in]{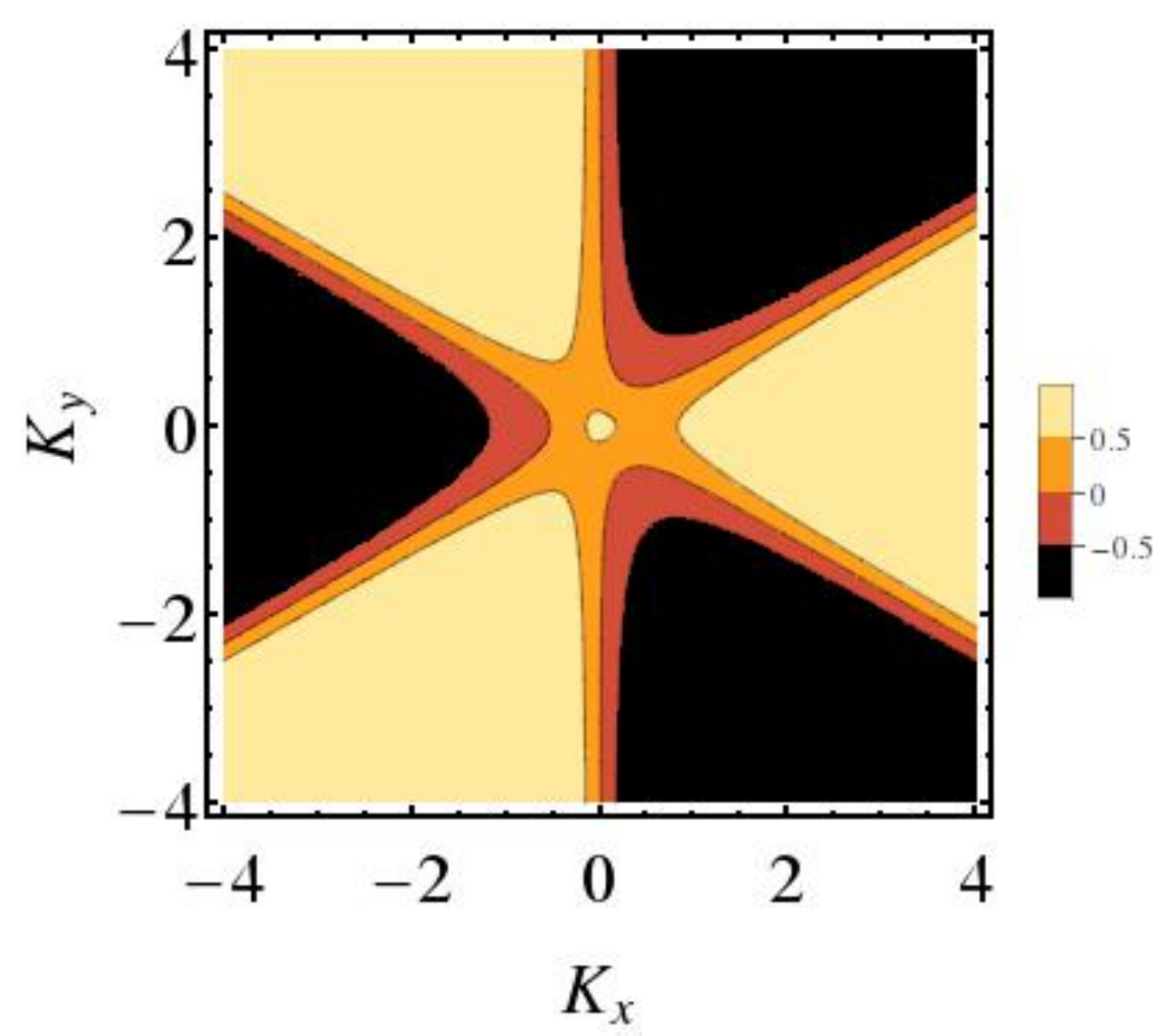}}
\caption{The $z$ component of spin ($S_{z}$ in units of $\hbar/2$) as a function of $k_{x}$, $k_{y}$ in units of nm$^{-1}$. The left figure shows $S_{z}$ of a normal gapless TI with finite warping. The parameters are $\Delta_{\omega}=0$, $\lambda=0.2$ eV nm$^{3}$, $a=0$. The figures shows that it has symmetric positive and negative regions. The right figure shows  $S_{z}$ of the Floquet warped TI surface. The parameters are $\Delta_{\omega}=0.03$ ev, $\lambda=0.2$ eV nm$^{3}$, $a=0.17$ nm.}
\label{spin3L}
\end{figure}

\section{In-Plane Spin Density} We now turn to a discussion of the total spin density. From Eq.(\ref{in-planex}) and Eq.(\ref{in-planey}), we calculate the total in-plane spin density (in units of $\hbar/2$) 
\begin{eqnarray}
S_{tot}(k,\theta)=\frac{\hbar v k\sqrt{1+a^2k^2-2ak\cos(3\theta)}}{\sqrt{\hbar^2 v^2 k^2(1+a^2k^2-2ak\cos(3\theta))+\Delta^2(k,\theta)}}\nonumber\\
\label{tot-spin}
\end{eqnarray}
In the absence of the polarized field (i.e., $a=0$ and $\Delta_{\omega}=0$), Eq.(\ref{tot-spin}) becomes,
\begin{eqnarray}
S_{tot}(k,\theta)=\frac{\hbar v k}{\sqrt{\hbar^2v^2k^2+\lambda^2k^6\cos^2(3\theta)}}
\label{tot-spin-abs}
\end{eqnarray}
The in-plane spin density in Eq.(\ref{tot-spin-abs}) of a time reversal invariant hexagonal warped TI has six-fold symmetry. It reaches a maximum value of $\hbar/2$ at $k=0$ independent of $\theta$, because the $k^6$ term in the denominator has negligible contribution. For other value of $k$, it follows the $\cos(3\theta)$ term and reaches $\hbar/2$ for $\theta=(2n+1)\pi/6$ ($n \in \mathbb{Z}$). The results are shown in the right panel of Fig.(\ref{spin-den})(See also colour density in Fig.(\ref{spin-plot})). The symmetry group reduces to $C_{3v}$ in Eq.(\ref{tot-spin}). $S_{tot}$ in Eq.(\ref{tot-spin}) takes zero values at $k=0$ or when $\cos(3\theta)=(1+a^2k^2)/2ak$ (for $k \neq 0$). The second condition demonstrates that it is possible to control spin density by tuning the parameter $a$. The results are shown in left panel of Fig.(\ref{spin-den}) (See also colour density in Fig.(\ref{spin3l-hg}(a)).

\section{Out-Of Plane Spin Component}
Here, We will discuss about out of-plane spin component ($S_{z}$). First, we consider the case without polarized optical field i.e. $a=0$ and $\Delta_{\omega}=0$. In this condition, the Eq.(\ref{out-of-plane}) becomes (also see Ref\cite{Li-PRB14})
\begin{eqnarray}
S_{z}=\pm \frac{\hbar}{2}\frac{\lambda (k^3_{x}-3k_{x}k^2_{y})}{\sqrt{\hbar^2v^2k^2+\lambda^2(k^3_{x}-3k_{x}k^2_{y})^2}}
\label{out-of-plane-wl}
\end{eqnarray}
From Eq.(\ref{out-of-plane-wl}), it is clear that, at a finite value of $\lambda$, $S_{z}$ is zero for $\theta=\pm \pi/6$ and $\pi/2$. For other value of $\theta$, it has symmetric positive and negative regions. The total average value of $S_{z}$ is zero due to time reversal invariance i.e, no net magnetization. The result is shown in left panel of Fig.(\ref{spin3L}). With the introduction of polarised field, the expression of $S_{z}$ is given in Eq.(\ref{out-of-plane}). Due to time reversal breaking, $S_{z}$ picks up maximum value of $\hbar/2$ at $k=0$. For other values of $k$, it has similar structure as earlier. The result is shown in right panel of Fig.(\ref{spin3L}). Note that, the out-of-plane spin component ($S_{z}$) belong to the same rotational symmetry group $C_{3v}$ for both time reversal and time reversal breaking system. But, as the time reversal breaking system becomes gapped at $k=0$, it picks up a finite value. 

\section{Angle of Deviation}
\begin{figure}
\rotatebox{0}{\includegraphics[width=0.96\columnwidth]{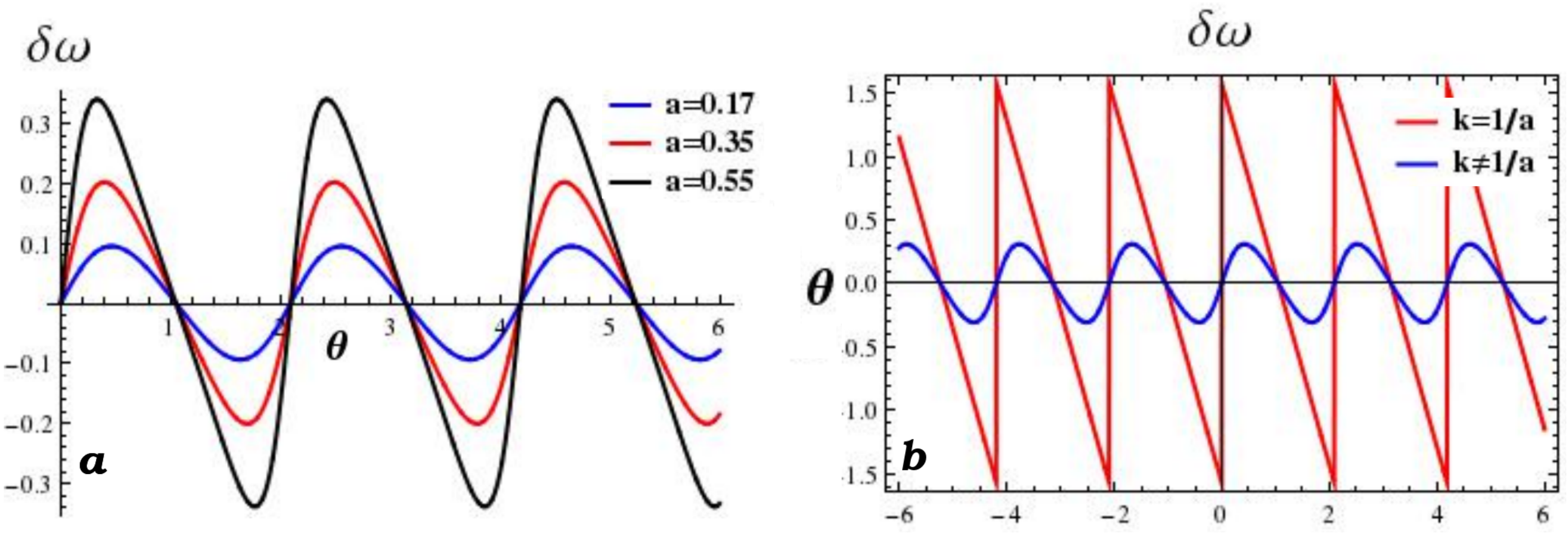}}
%\rotatebox{0}{\includegraphics[width=1.7in]{delta.pdf}}
%\rotatebox{0}{\includegraphics[width=3.0in]{delta-a.pdf}}
\caption{Plot of angle of deviation ($\delta_{\omega}$) with $\theta$. (a) Shows $\delta_{\omega}$ as a function of $\theta$ for different values of $a$ with $k=0.55$ nm$^{-1}$.(b) Shows $\delta_{\omega}$ as a function of $\theta$ for $k=\frac{1}{a}$ and $k\neq \frac{1}{a}$. $\delta_{\omega}$ is undefined at $\theta=2n\pi/3$ ($n \in \mathbb{Z}$) for $k=\frac{1}{a}$ since total in-plane density $S_{tot}$ becomes zero (see also Fig.(\ref{spin-den})).}
\label{dev-op}
\end{figure}

We will end our discussion by calculating angle of deviation($\delta_{\omega}$) between in-plane spin and momentum of the electron. It will help us understand how the in-plane spin components are oriented along momentum direction. From Eq.(\ref{in-planex}) and Eq.(\ref{in-planey}), one can calculate $\delta_{\omega}$:
\begin{eqnarray}
\delta_{\omega}=\cos^{-1}[\frac{-ak\sin(3\theta)}{\sqrt{1+a^2k^2-2ak\cos(3\theta)}}]-\frac{\pi}{2}
\label{deviation}
\end{eqnarray}
So, the angle of deviation $\delta_{\omega}$ is function of $k$ and also of $\theta$. From Eq.(\ref{deviation}), it is clear that the angle of deviation is zero for $\theta=0$ or $\pm \pi/3$, independent of $k$. For other values of $\theta$, $\delta_{\omega}$ has symmetric positive and negative regions.  The result is shown in Fig(\ref{dev-op}). As we have discussed earlier the higher order warping term also breaks the spin momentum locking on the surface of the Dirac cone and it was shown that $\delta_{\omega}$ is zero along two symmetry directions ($\Gamma-K$ and $\Gamma-M$). But in our case, it is only zero along symmetry line $\Gamma-K$ and $\theta=\pi/3$. This is due to fact that in our case, off-resonant light has asymmetrical coupling to the electron momentum. For example, the form of the Hamiltonian along $\Gamma-M$ direction is given by,
\begin{eqnarray}
\mathcal{H}^{\Gamma-M}_{eff}=\hbar v[ak^2_{y}\sigma_{y}-k_{y}\sigma_{x}]+\Delta_{\omega}\sigma_{z}
\label{gamma-m}.
\end{eqnarray}
The term $ak^2_{y}\sigma_{y}$ in Eq.(\ref{gamma-m}) breaks the mirror symmetry resulting in a finite $S_{y}$ value. So, in the spin structure, both the in-plane spin ($S_{x}$ and $S_{y}$) has a finite value and the resultant spin component becomes non-orthogonal to the $k_{y}$-axis. Similarly, the form of the Hamiltonian along $\Gamma-K$ direction is given by,
\begin{eqnarray}
\mathcal{H}^{\Gamma-K}_{eff}=\hbar v(k_{x}-ak^2_{x})\sigma_{y}+(\lambda k^3_{x}+\Delta_{\omega})\sigma_{z}.
\label{gamma-k}
\end{eqnarray}
So, only the $S_{y}$ spin component exists and remains orthogonal to the $k_{x}$-axis. (Similarly, one can see the spin-momentum orthogonality/non-orthogonality at other points in BZ also).Now we take the higher order warping term $\mathcal{H}_{hw}=i\xi(k^5_{+}\sigma_{+}-k^5_{-}\sigma_{-})$ as a perturbation to see whether it changes the $\delta_{\omega}$. Along $\Gamma-M$ and $\Gamma-K$ direction the warping term modifies the Hamiltonian of Eq.(\ref{gamma-m}) and Eq.(\ref{gamma-k}) by $\mathcal{H}^{\Gamma-M}_{hw}=-2\xi k^5_{y}\sigma_{x}$ and $\mathcal{H}^{\Gamma-K}_{hw}=-2\xi k^5_{x}\sigma_{y}$. So, the behaviour of $\delta_{\omega}$ still remains the same (It is also invariant for $\theta=\pi/3$ since $\mathcal{H}_{hw}$ belongs to same rotational group of $\mathcal{H}_{eff}$). In Fig.(\ref{dev-op})(a), we take different values of $a$ and it is seen that the $\delta_{\omega}$ increases with the value of $a$. Eq.(\ref{deviation}) also tells that for $k=1/a$, $\delta_{\omega}$ is undefined at $\theta=2n\pi/3$ ($n \in \mathbb{Z}$). Actually, at that point the in-plane density is zero (see also Fig.(\ref{spin-den})) and so the $\delta_{\omega}$ is meaningless. We show our result in Fig(\ref{dev-op})(b) for $k=1/a$ and $k \neq 1/a$. The results shows that it is possible to tune $\delta_{\omega}$ at a fixed value of $k$ in BZ by tuning the parameter $a$.

\section{Conclusions}
We have shown that the locking of spin with momentum is broken when a TI surface is irradiated with an off-resonant light field, even when this symmetry is not broken explicitly by the underlying time-reversal-symmetric Hamiltonian. We carry out the discussions in the context of a model Hamiltonian which has hexagonal warping. In this case, a novel spin-texture emerges, consistent with the $C_{3v}$ symmetry of the underlying Hamiltonian. Since ultimately the spin-texture has to be consistent with the symmetries of the system, the spin-texture we obtain is distinctly different from the one in which time-reversal symmetry of the system is broken using magnetic doping. The spin texture we obtain by explicitly breaking time-reversal symmetry is also less symmetric than the one which would be obtained if one included a further fifth order in momentum term in the Hamiltonian consistent with $C_{3v}$ symmetry and preserving time-reversal symmetry.\\
Since irradiation by polarised light beam is experimentally easily accessible, our work presents the simplest way of experimentally realising and controlling spin-texture. We demonstrate theoretically how the in-plane spin density, out-of-plane spin component and angle of deviation can be tuned by tuning the strength of the polarised optical field. Since higher order terms in the Hamiltonian do not break any symmetry not already broken by the polarised electromagnetic field, our results will remain unchanged under inclusion of all such terms. Finally, our work can also be extended to TIs with different lattice symmetries. Generically, the spin-texture in all TIs under irradiation will only have the symmetries of the underlying lattice. We hope that our contribution will help in understanding and study of spin-textures in newer TIs that are being continually discovered.
%\section{Acknowledgement}
%I am immensely indebeted to S. R. Hassan for many stimulating discussions. Also a word of thank to S. Sengupta, V. Lukose, T. Das and A. Maitra for useful correspondense.

\section{Appendix}
Here we present a brief discussion on the Floquet formalism and show that the effective static model can equivalently be obtained using the Floquet theory. Under irradiation, the Hamiltonian $\mathcal{H}(t)=\mathcal{H}_{0}+\mathcal{V}(t)$ becomes time periodic i.e., $\mathcal{H}(\vec{k},t+T)=\mathcal{H}(\vec{k},t)$ with $T=2\pi/\omega$. Due to the discrete time peridicity, the solution of Schr\"odinger equation,
\begin{eqnarray} 
i\hbar\partial_{t}\ket{\Psi(t)}=\mathcal{H}(\vec{k},t)\ket{\Psi(t)}
\label{Sch}
\end{eqnarray}
has the form
\begin{eqnarray}
\ket{\Psi(t)}=e^{-i\mathcal{E}t/\hbar}\ket{\Phi(t)}
\end{eqnarray}
which is a product of the phase factor involving the Floquet quasienergy $\mathcal{E}$ and the time-periodic wave function $\ket{\Phi(t+T)}=\ket{\Phi(t)}$. Time periodicity guarantees that we can use a discrete Fourier transformation,
\begin{eqnarray}
\mathcal{V}(t)&=&\sum_{m}e^{-im\omega\hbar t}H_{m}\nonumber\\
\ket{\Phi(t)}&=&\sum_{m}e^{-im\omega\hbar t}\ket{\Phi_{m}}
\label{Fourier}
\end{eqnarray}
When Eq.(\ref{Fourier}) substituted in Eq.(\ref{Sch}), the eigenvalue equation becomes matrix form as,
\begin{eqnarray}
\sum_{m}(H_{n-m}-m\hbar \omega \delta_{mn})\ket{\Phi_{m}}=\mathcal{E}\ket{\Phi_{m}}
\label{time-indp}
\end{eqnarray}
i.e., the matrix form of parenthesis in Eq.(\ref{time-indp})can be written in tridiagonal form as,
%Here we present a brief discussion on the derivation of effective Hamiltonian within the off resonant approximation for the periodic Floquet Hamiltonian $\mathcal{H}(t)=\mathcal{H}_{0}+V(t)$, where $\mathcal{H}_{0}$ is the static contribution and $V(t+T)=V(t)$ is the time periodic interaction. Transforming to the Fourier space we get the Floquet Hamiltonian for a monochromatic perturbation in matrix form as \cite{Medina,John-PRB15}
\begin{eqnarray*}
\begin{pmatrix}
....\\
... & V_{-1} & \mathcal{H}_{-2} & V_{+1} & 0 & 0 & 0 & ...\\
... & 0 & V_{-1} & \mathcal{H}_{-1} & V_{+1} & 0 & 0 & ...\\
... & 0 & 0 & V_{-1} & \mathcal{H}_{0} & V_{+1}& 0 &...\\
... & 0 & 0 & 0 & 0 & V_{-1} & \mathcal{H}_{1} & V_{+1}...\\
....& 0 & 0 & 0 & 0 & 0 & V_{-1} & \mathcal{H}_{2}&.....\\
\end{pmatrix}
\end{eqnarray*}
with
\begin{eqnarray}
V_{m}=\frac{1}{T}\int^{T}_{0} dt \mathcal{H}(t) e^{-im\hbar \omega t}
\end{eqnarray}
define over a period and $\mathcal{H}_{m}=\mathcal{H}_{0}+m\hbar\omega$. The term with phase factor $e^{-im\hbar \omega t}$ in the Hamiltonian induces a transition from Floquet mode $m$ to $m \pm 1$. The eigenvalue of time independent in equation (\ref{time-indp}) is given for any general value of $m$ as,
\begin{eqnarray}
\phi=\begin{pmatrix}
\phi_{-m}\\
.\\
\phi_{-1}\\
\phi_{0}\\
\phi_{1}\\
.\\
\phi_{m-1}\\
\phi_{m}
\end{pmatrix}
\end{eqnarray}

In general $m$ varies from $-\infty$ to $\infty$. But the matrix form can be reduced if the frequency of polarised field is high enough. Here we will neglect any interband and intraband electron transition i.e. frequency is off-resonant. Such off-resonant condition is satisfied for the frequency $\hbar\omega \gg ||\mathcal{H}_{0}||$. Under the condition, we restrict our formalism for low values of $m$ ($m=-1,0,1$). This leads three coupled equations \cite{Medina,John-PRB15}
\begin{eqnarray}
\mathcal{H}_{-1}\phi_{-1}+V_{+1}\phi_{0}&=&\mathcal{E}\phi_{-1}\\
\label{second-eq}
V_{-1}\phi_{-1}+\mathcal{H}_{0}\phi_{0}+V_{+1}\phi_{+1}&=&\mathcal{E} \phi_{0}\\
\mathcal{H}_{+1}\phi_{+1}+V_{-1}\phi_{0}&=&\mathcal{E}\phi_{+1}
\end{eqnarray}
From the first and last equations we get,
\begin{eqnarray}
\phi_{-1}&=&(\mathcal{E}-\mathcal{H}_{-1})^{-1}V_{+1}\phi_{0}\\
\phi_{+1}&=&(\mathcal{E}-\mathcal{H}_{+1})^{-1}V_{-1}\phi_{0}
\end{eqnarray}
Putting $\phi_{-1}$ and $\phi_{+1}$ back into Eq.(\ref{second-eq}), one can get an approximate solution as,
\begin{eqnarray}
(\mathcal{H}_{0}+\frac{V_{-1}V_{+1}}{\hbar\omega}-\frac{V_{+1}V_{-1}}{\hbar \omega})\phi_{0}\simeq \mathcal{E} \phi_{0}
\label{floq-eff}
\end{eqnarray}
i.e.,
\begin{eqnarray}
\mathcal{H}_{eff}\phi_{0}\simeq \mathcal{E}\phi_{0}
\end{eqnarray}
where, the effective Hamiltonian given as,
\begin{eqnarray}
\mathcal{H}_{eff}\simeq \mathcal{H}_{0}+\frac{[V_{-1},V_{+1}]}{\hbar \omega}
\end{eqnarray}
%For any value $n$, following this method it is easy to show the effective Hamiltonian becomes
%\begin{eqnarray}
%\mathcal{H}_{F}\simeq \mathcal{H}_{0}+\sum^{\infty}_{n=1}(-1)^{n-1}\frac{1}{n\hbar \omega}\frac{[V^{n}_{-1},V^{n}_{1}]}{[(n-1)!\hbar \omega]^{2(n-1)}}
%\end{eqnarray}
So, a static effective Hamiltonian is possible to find in the Floquet regime under the off-resonant condition. The second and third term in Eq.(\ref{floq-eff}) are associated with a virtual photon emission and absorption process\cite{Demler-PRB11}.

\acknowledgments
I am indebeted to S. R. Hassan for many stimulating discussions. I acknowledge A. Maitra for helping me to improve the revised version. I am benifited by comments and suggestions from two anonymous referees. I also thank S. Sengupta, T. Das, A. Dutta and A. Laskar for useful correspondense.


\begin{thebibliography}{0}

\bibitem{Zhang-08} M. K\"{o}nig, H. Buhmann, L. W. Molenkamp, T. Haughes, C.-X. Liu, X.-L. Qi, and S.-C. Zhang, Journal of the Physical Society of Japan 77, 031007 (2008) 
\bibitem{Zhou-08} B. Zhou, H.-Z. Lu, R.-L. Chu, S.-Q. Shen and Q. Niu, Phys. Rev. Lett. 101, 246807 (2008)
\bibitem{kane-rmp} Colloquium: Toplogical insulators, Rev. Mod. Phys, 82, 3045 (2010)
\bibitem{Hughes-Nat08} X.-L. Qi, T. L. Hughes and S.-C. Zhang, Nat Phys 4, 273 (2008)
\bibitem{Zhang-Nat09} H. Zhang, C.-X. Liu, X. Dai, Z. Fang and S.-C. Zhang, Nature Phys. 5, 438 (2009)
\bibitem{Niu-Nat10}Y. Zhang et. al, Nature Phys. 6, 584 (2010)
\bibitem{Chadov-NAM10} S. Chadov, X.-L. Qi, J. Kubler, G. H. Fecher, and C. F. S.-C. Zhang, Nat. Mater.9, 546 (2010)
\bibitem{Hsieh-Nat09} D. Hsieh et. al, Nature (London) 460, 1101 (2009)
\bibitem{Hsieh-PRL09} D. Hsieh et al., Phys. Rev. Lett. 103, 146401 (2009)
%\bibitem{Hsieh-Nat08} D. Hsieh, D. Qian, L. Wray, Y. Xia, Y. S. Hor, R. J. Cava, and M. Z. Hasan, Nature (London) 452, 970 (2008)
%\bibitem{Hsieh-Nat08} D. Hsieh, D. Qian, L. Wray, Y. Xia, Y. S. Hor, R. J. Cava, and M. Z. Hasan, Nature (London) 452, 970 (2008)
\bibitem{Lu-PRL11} H.-Z. Lu, J. R. Shi and S.-Q.Shen, Phys. Rev. Lett. 107, 076801 (2011)
\bibitem{Hikami-PTP80} S. Hikami, A. I. Larkin and Y. Nagaoka, Prog. Theor. Phys. 63, 707-710 (1980)
\bibitem{Bergmann-PR84} G. Bergmann, Phys. Rep. 107, 1-58 (1984)
\bibitem{Su-Science11} Su-Yang Xu et. al., Science 332:560, 2011
\bibitem{Hsieh-Nat08} D. Hsieh, D. Qian, L. Wray, Y. Xia, Y. S. Hor, R. J. Cava, and M. Z. Hasan, Nature (London) 452, 970 (2008)
%\bibitem{Hsieh-Nat09} D. Hsieh et. al, Nature (London) 460, 1101 (2009)
%\bibitem{Lu-PRL11} H.-Z. Lu, J. R. Shi and S.-Q.Shen, Phys. Rev. Lett. 107, 076801 (2011)
%\bibitem{Hikami-PTP80} S. Hikami, A. I. Larkin and Y. Nagaoka, Prog. Theor. Phys. 63, 707-710 (1980)
%\bibitem{Bergmann-PR84} G. Bergmann, Phys. Rep. 107, 1-58 (1984)
\bibitem{OV-PRL10} O. V. Yazyev, J. E. Moore and S. G. Louie, Phys. Rev. Lett. 105, 266806
\bibitem{Liu-Nature12} S.-Y. Xu et. al., Nature Physics, 8, 616 (2012)
\bibitem{Qin-Niu-PRL} Q. Niu, C.-X.Liu, C. Xu, X.-L. Qi and S.-C. Zhang, Phys. Rev. Lett, 102, 156603 (2009)
\bibitem{Fu} L. Fu, Phys. Rev. Lett.103,266801 (2009)
%\bibitem{Hsieh-PRL09} D. Hsieh et al., Phys. Rev. Lett. 103, 146401 (2009)
\bibitem{Chen-Scince10} Y. L. Chen et al., Science 329, 659 (2010)
\bibitem{Li-PRB13} Z. Li and J . P. Carbotte, Phys. Rev. B 87, 155416 (2013)
\bibitem{Li-PRB14} Z. Li and J. P. Carbotte, Phys. Rev. B 89, 165420 (2014)
%\bibitem{berche-epj} D. Sinha and B. Berche, Eur. Phys. J. B, 89: 57 (2016)
\bibitem{Bansil-PRB11} S. Basak et. al, Phys. Rev. B, 84, 121401 (R) (2011) 
\bibitem{Demler-PRB11} T. Kitagawa, T. Oka, A. Brataas, L. Fu, and E. Demler, Phys. Rev. B 84, 235108 (2011)
\bibitem{Cayssol-PSS13} J. Cayssol, B. Dora, F. Simon and R. Moessner, Phys. Status Solidi RRL 7, 101 (2013)
\bibitem{Podo-PRB13} Y. T. Katan and D. Podolosky, Phys. Rev. B, 88, 224106 (2013)
\bibitem{Lago-PRA15} V. D. Lago, M. Atala and L. E. F. Foa Torres, Phys. Rev. A, 92, 023624 (2015)
\bibitem{Calvo-PRB15} H. L. Calvo, L. E. F. Foa Torres, P. M. Perez-Piskunow, C. A. Balseiro and G. Usaj, Phys. Rev. B, 91, 241404 (R) (2015)
\bibitem{Jun-PRL10} J. Inoue and A. Tanaka, Phys. Rev. Lett, 105, 017401 (2010)
\bibitem{Dora-PRL12} B. Dora, J. Cayssol, F. Simon, and R. Moessner, Phys. Rev. Lett, 108, 056602 (2012)
\bibitem{Lovey-ar} D. A. Lovey, G. Usaj, L. E. F. Foa Torres and C. A. Balseiro, Phys. Rev. B, 93, 245434 (2016)
\bibitem{Zhai-PRB14} X. Zhai and G. Zin, Phys. Rev. B, 89, 235416 (2014)
\bibitem{Zhou-PRB15} X. Zhou, Y. Xu and G. Jin, Phys. Rev. B, 92, 235436 (2016)
\bibitem{Tahir-PRB} M. Tahir and P. Vasilopoulos, Phys. Rev. B, 91, 115311 (2015)
\bibitem{weyl-semi} H. H\"ubener et. al., "Creating stable Floquet-Weyl semimetals by laser driving of 3D Dirac materials", arXiv:1604.03399
\bibitem{Wang-Science13} Y. H. Wang, H. Steinberg, P. Jarillo-Herrero, and N. Gedik, Science 342, 453 (2013)
\bibitem{Rechtsman-Nat13} M. C. Rechtsman et. al, Nature (London) 496, 196 (2013)
\bibitem{Ezawa-PRL13} M. Ezawa, Phys. Rev. Lett, 110, 026603 (2013)
\bibitem{Xin-CHIN14} Z. H.-Xin, W. T.-Tong, G. J.-Song, L. Shuai, S. Y.-Jun, L. G.-Lin, CHIN. PHYS. LETT, Vol31, No.3 (2014) 030503
%\bibitem{Zhai-PRB14} X. Zhai and G. Zin, Phys. Rev. B, 89, 235416 (2014)
%\bibitem{Zhou-PRB15} X. Zhou, Y. Xu and G. Jin, Phys. Rev. B, 92, 235436 (2016)
%\bibitem{Lovey-ar} D. A. Lovey, G. Usaj, L. E. F. Foa Torres and C. A. Balseiro, "Floquet bound state around defects and adatoms in graphene", arXiv:1603.04398
%\bibitem{Tahir-PRB} M. Tahir and P. Vasilopoulos, Phys. Rev. B, 91, 115311 (2015)
\bibitem{berche-epj} D. Sinha and B. Berche, Eur. Phys. J. B, 89: 57 (2016)
%\revision{\bibitem{Qin-Niu-PRL} Q. Niu, C.-X.Liu, C. Xu, X.-L. Qi and S.-C. Zhang, Phys. Rev. Lett, 102, 156603 (2009)}
\bibitem{Medina} E. Medina and H. Pastawski, Rev. Mex. Phys. 47, 1 (2001)
\bibitem{John-PRB15} A. Lopez, A. Scholz, B. Santos and J. Schliemann, Phys. Rev. B 91, 125105 (2015)
\end{thebibliography}
\end{document}